# Bounds on the Decoding Complexity of Punctured Codes on Graphs

| Henry D. Pfister | Igal Sason | Rüdiger Urbanke |
| --- | --- | --- |
| Qualcomm, Inc. | Technion | EPFL |
| CA 92121, USA | Haifa 32000, Israel | Lausanne 1015, Switzerland |
| hpfister@qualcomm.com | Sason@ee.technion.ac.il | Rudiger.Urbanke@epfl.ch |

**Abstract**

We present two sequences of ensembles of non-systematic irregular repeat-accumulate codes which asymptotically (as their block length tends to infinity) achieve capacity on the binary erasure channel (BEC) with *bounded complexity* per information bit. This is in contrast to all previous constructions of capacity-achieving sequences of ensembles whose complexity grows at least like the log of the inverse of the gap (in rate) to capacity. The new bounded complexity result is achieved by puncturing bits, and allowing in this way a sufficient number of state nodes in the Tanner graph representing the codes. We also derive an information-theoretic lower bound on the decoding complexity of randomly punctured codes on graphs. The bound holds for every memoryless binary-input output-symmetric channel, and is refined for the BEC.

## 1 Introduction

During the last decade, there have been many exciting developments in the construction of low-complexity error-correction codes which closely approach the capacity of many standard communication channels with feasible complexity. These codes are understood to be codes defined on graphs, together with the associated iterative decoding algorithms. By now, there is a large collection of these codes that approach the channel capacity quite closely with moderate complexity.

The first capacity-achieving sequences of ensembles of low-density parity-check (LDPC) codes for the binary erasure channel (BEC) were found by Luby et al. [3, 4] and Shokrollahi [11]. Following these pioneering works, Oswald and Shokrollahi presented in [5] a systematic study of capacity-achieving degree distributions (d.d.) for sequences of ensembles of LDPC codes whose transmission takes place over the BEC. Capacity-achieving ensembles of irregular repeat-accumulate (IRA) codes for the BEC were introduced and analyzed in [1, 10], and also capacity-achieving ensembles for erasure channels with memory were designed and analyzed in [6].

In [2], Khandekar and McEliece discussed the complexity of achieving the channel capacity on the BEC, and more general channels with vanishing bit error probability. They conjectured that if the achievable rate under message-passing iterative (MPI) decoding is a fraction $1-\varepsilon$ of the channel capacity, then for a wide class of channels, the encoding complexity scales like $\ln\frac{1}{\varepsilon}$ and the decoding complexity scales like $\frac{1}{\varepsilon}\ln\frac{1}{\varepsilon}$. This conjecture is based on the assumption that the number of edges (per information bit) in the associated bipartite graph scales like $\ln\frac{1}{\varepsilon}$, and the required number of iterations under

MPI decoding scales like $\frac{1}{\varepsilon}$. However, for codes defined on graphs which are transmitted over a BEC, the decoding complexity under the MPI algorithm behaves like $\ln \frac{1}{\varepsilon}$ (same as encoding complexity) [3, 9, 11]. This is since the absolute reliability provided by the BEC allows every edge in the graph to be used only once during MPI decoding.

In [9], Sason and Urbanke considered the question of how sparse can parity-check matrices of binary linear codes be, as a function of their gap (in rate) to capacity (where this gap depends on the channel and the decoding algorithm). If the code is represented by a standard Tanner graph without state nodes, the decoding complexity under MPI decoding is strongly linked to the density of the corresponding parity-check matrix (i.e., the number of edges in the graph per information bit). In particular, they considered an arbitrary sequence of binary linear codes which achieves a fraction $1-\varepsilon$ of the capacity of a memoryless binary-input output-symmetric (MBIOS) channel with vanishing bit error probability. By information-theoretic tools, they proved that for every such sequence of codes and every sequence of parity-check matrices which represent these codes, the asymptotic density of the parity-check matrices grows at least like $\frac{K_1+K_2 \ln \frac{1}{\varepsilon}}{1-\varepsilon}$ where $K_1$ and $K_2$ are constants which were given explicitly as a function of the channel statistics (see [9, Theorem 2.1]). It is important to mention that this bound is valid under ML decoding, and hence, it also holds for every sub-optimal decoding algorithm. The tightness of the lower bound for MPI decoding on the BEC was demonstrated in [9, Theorem 2.3] by analyzing the capacity-achieving sequence of check-regular LDPC-code ensembles introduced by Shokrollahi [11]. Based on the discussion in [9], it follows that for every iterative decoder which is based on the standard Tanner graph, there exists a fundamental tradeoff between performance and complexity, and the complexity (per information bit) becomes *unbounded* when the gap between the achievable rate and the channel capacity vanishes. Therefore, it was suggested in [9] to study if better tradeoffs can be achieved by allowing more complicated graphical models (e.g., graphs which also involve state nodes).

In this paper, we present sequences of capacity-achieving ensembles for the BEC with bounded complexity under MPI decoding. The new ensembles are non-systematic IRA codes with properly chosen d.d. (for background on IRA codes, see [1] and Section 2). The new bounded complexity results improve on the results in [10], and demonstrate the superiority of properly designed non-systematic IRA codes over systematic IRA codes (since with probability 1, the complexity of any sequence of ensembles of systematic IRA codes becomes *unbounded* under MPI decoding when the gap between the achievable rate and the capacity vanishes [10, Theorem 1]). The new bounded complexity result is achieved by allowing a sufficient number of state nodes in the Tanner graph representing the codes. Hence, it answers in the affirmative a fundamental question which was posed in [9] regarding the impact of state nodes in the graph on the performance versus complexity tradeoff under MPI decoding. We suggest a particular sequence of capacity-achieving ensembles of non-systematic IRA codes where the degree of the parity-check nodes is 5, so the complexity per information bit under MPI decoding is equal to $\frac{5}{1-p}$ when the gap (in rate) to capacity vanishes ($p$ designates the bit erasure probability of the BEC). Computer simulation results for these ensembles appear to agree with this analytical result. It is worth noting that our method of truncating the check d.d. is similar to the bi-regular check d.d. introduced in [13] for non-systematic IRA codes.

We also present in this paper an information-theoretic lower bound on the decoding complexity of randomly punctured codes on graphs. The bound holds for every MBIOS channel with a refinement for the particular case of a BEC.

The structure of the paper is as follows: Section 2 provides preliminary material on ensembles of IRA codes, Section 3 presents our main results. For the sake of brevity, we omit the proofs; the interested reader is referred to [7] for the proofs, and also for other analytical and numerical results w.r.t. the considered d.d. which are discussed in the full paper version [7]. Practical considerations and simulation results for our ensembles of IRA codes are presented in Section 4. We conclude our discussion in Section 5.

## 2  IRA Codes

We consider in our first two theorems ensembles of non-systematic IRA codes. We assume that all information bits are punctured. The Tanner graph of these codes is shown in Fig. 1. These codes can be viewed as serially concatenated codes, where the outer code is a mixture of repetition codes of varying order and the inner code is generated by a differential encoder with puncturing. We define these ensembles by a uniform choice of the interleaver separating the component codes.

We assume that a randomly selected code from this ensemble is used to communicate over a BEC with erasure probability $p$. The asymptotic performance of the MPI decoder (as the block length tends to infinity) can be analyzed by tracking the average fraction of erasure messages which are passed in the graph of Fig. 1 during the $l^{\text{th}}$ iteration. This technique is known as density evolution (DE) [8]. Using the same notation as in [1], let $x_0^{(l)}$ be the probability of erasure for a message from information nodes to parity-check nodes, $x_1^{(l)}$ be the probability of erasure from parity-check nodes to code nodes, $x_2^{(l)}$ be the probability of erasure from code nodes to parity-check nodes, and finally, let $x_3^{(l)}$ be the probability of erasure for messages from parity-check nodes to information nodes (see Fig. 1). We now assume that we are at a fixed point of the MPI decoding algorithm, and solve for $x_0$. We obtain the following equations:

$$x_1 = 1 - (1 - x_2)R(1 - x_0), \tag{1}$$
$$x_2 = px_1, \tag{2}$$
$$x_3 = 1 - (1 - x_2)^2 \rho(1 - x_0), \tag{3}$$
$$x_0 = \lambda(x_3), \tag{4}$$

where $R(x) = \sum_{i=1}^{\infty} R_i x^i$ is a power series where the coefficient $R_i$ denotes the fraction of parity-check nodes that are connected to $i$ information nodes (as is depicted in Fig. 1, we note that every parity-check node is also connected to two code bits; this is a result of the differential encoder which is the inner code of the considered construction of serially concatenated and interleaved codes). The power series $\rho(x) = \sum_{i=1}^{\infty} \rho_i x^{i-1}$ is given from the edge perspective, where $\rho_i$ designates the fraction of edges in the Tanner graph (see Fig. 1) which are connected to parity-check nodes of degree $i$; as before, the degree of the parity-check nodes refers to the number of edges which are only connected to information nodes. Therefore, $R(\cdot)$ and $\rho(\cdot)$ are related by the equation

$$R(x) = \frac{\int_0^x \rho(t) \, dt}{\int_0^1 \rho(t) \, dt}.$$

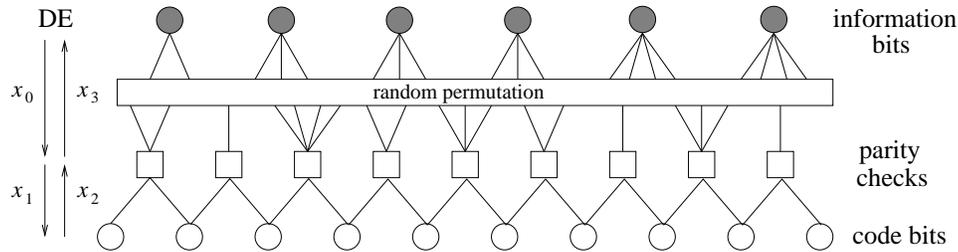

Figure 1: The Tanner graph of IRA codes.

The only difference between equations (1)–(4) and those in [1] is the removal of a $p$ in (4), since all the information bits are punctured in the considered ensemble of codes. Solving this set of equations for a fixed point of MPI decoding provides the equation

$$x_0 = \lambda \left( 1 - \left[ \frac{1-p}{1 - pR(1-x_0)} \right]^2 \rho(1-x_0) \right). \tag{5}$$

If Eq. (5) has no solution in the interval $(0, 1]$ and the RHS is less than one at $x_0 = 1$, then DE analysis of MPI decoding must converge to a bit erasure probability of zero. Therefore, the condition that

$$\lambda \left( 1 - \left[ \frac{1-p}{1 - pR(1-x)} \right]^2 \rho(1-x) \right) < x, \quad \forall x \in (0, 1], \tag{6}$$

implies that MPI decoding obtains a vanishing bit erasure probability as the block length tends to infinity.

The design rate of the ensemble of non-systematic IRA codes can be computed by matching edges in decoding graph shown in Fig. 1. This implies that the design rate is equal to

$$R^{\text{IRA}} = \frac{\int_0^1 \lambda(x) \, dx}{\int_0^1 \rho(x) \, dx}. \tag{7}$$

Furthermore, we show in [7] that $R^{\text{IRA}} = 1 - p$ for any pair of d.d. $(\lambda, \rho)$ which satisfies Eq. (5) for all $x_0 \in [0, 1]$.

In order to find a capacity-achieving ensemble of IRA codes, we generally start by finding a pair of d.d. $(\lambda, \rho)$ with non-negative power series expansions which also satisfies Eq. (5) for all $x_0 \in [0, 1]$. Next, we slightly modify $\lambda(\cdot)$ or $\rho(\cdot)$ so that Eq. (6) is satisfied and the new design rate in Eq. (7) is equal to $(1 - \varepsilon)(1 - p)$ for an arbitrarily small $\varepsilon$. Since the capacity of the BEC is $1 - p$, this gives an ensemble which has vanishing bit erasure probability under MPI decoding at rates which are arbitrarily close to capacity.

## 3 Main Results

**Definition 1.** Let $\{\mathcal{C}_m\}$ be a sequence of binary linear codes of rate $R_m$, and assume that for every $m$, the codewords of the code $\mathcal{C}_m$ are transmitted with equal probability over a channel whose capacity is $C$. This sequence is said to *achieve a fraction $1 - \varepsilon$ of the channel capacity with vanishing bit error probability* if $\lim_{m \to \infty} R_m \geq (1 - \varepsilon)C$, and there exists a decoding algorithm under which the average bit error probability of the code $\mathcal{C}_m$ tends to zero in the limit where $m$ tends to infinity.[1]

---
[1] We refer to vanishing bit erasure probability for the particular case of a BEC.

**Definition 2.** Let $\mathcal{C}$ be an ensemble of LDPC or IRA codes whose d.d., $\lambda(\cdot)$ and $\rho(\cdot)$, can be chosen arbitrarily (subject to possibly some constraints). The *encoding and the decoding complexity* are measured in operations per information bit which are required for achieving a fraction $1-\varepsilon$ of the channel capacity with vanishing bit erasure probability. Unless the pair of d.d. is specified, the encoding and the decoding complexity are measured with respect to *the best ensemble* (i.e., for the optimized pair of d.d.), and refer to the average complexity over this ensemble (as the block length of the codes tends to infinity, the complexity of a typical code from this ensemble concentrates to the average complexity). We denote the encoding and the decoding complexity by $\chi_E(\varepsilon, \mathcal{C})$ and $\chi_D(\varepsilon, \mathcal{C})$, respectively.

We note that for the BEC, both the encoding and the decoding complexity of IRA codes under MPI decoding is equal to the normalized number of edges per information bit in the associated Tanner graph.

**Theorem 1 (Bit-Regular Ensembles with Bounded Complexity).** Consider the ensemble of bit-regular non-systematic IRA codes $\mathcal{C}$, where the d.d. of the information bits is given by
$$\lambda(x) = x^{q-1}, \quad q \geq 3$$
which implies that each information bit is repeated $q$ times. Assume that the transmission takes place over a BEC with erasure probability $p$, and let the d.d. of the parity-check nodes[2] be
$$\rho(x) = \frac{1 - (1-x)^{\frac{1}{q-1}}}{\left[1 - p\left(1 - qx + (q-1)\left[1 - (1-x)^{\frac{q}{q-1}}\right]\right)\right]^2}.$$

Let $\rho_n$ be the coefficient of $x^{n-1}$ in the power series expansion of $\rho(x)$ and, for an arbitrary $\epsilon \in (0,1)$, define $M(\varepsilon)$ to be the smallest positive integer[3] $M$ such that
$$\sum_{n=M+1}^{\infty} \rho_n < \frac{\varepsilon}{q(1-p)}.$$

The $\varepsilon$-truncated d.d. of the parity-check nodes is now given by
$$\rho_\varepsilon(x) = \left(1 - \sum_{n=2}^{M(\varepsilon)} \rho_n\right) + \sum_{n=2}^{M(\varepsilon)} \rho_n x^{n-1}.$$

For $q=3$ and $p \in (0, \frac{1}{13}]$, the polynomial $\rho_\varepsilon(\cdot)$ has only non-negative coefficients, and the d.d. pair $(\lambda, \rho_\varepsilon)$ achieves a fraction $1-\varepsilon$ of the channel capacity with vanishing bit erasure probability under MPI decoding. Moreover, the complexity (per information bit) of encoding and decoding satisfies
$$\chi_E(\varepsilon, \mathcal{C}) = \chi_D(\varepsilon, \mathcal{C}) < q + \frac{2}{(1-p)(1-\varepsilon)}.$$

In the limit as $\varepsilon$ tends to zero, the capacity is achieved (using MPI decoding) with a *bounded* complexity of $q + \frac{2}{1-p}$.

---
[2] The d.d. of the parity-check nodes refers only to the connection of the parity-check nodes with the information nodes. Every parity-check node is also connected to *two code bits* (see Fig. 1), but this is not included in $\rho(x)$.

[3] The existence of $M(\varepsilon)$ for $\varepsilon \in (0,1)$ follows from the fact that $\rho_n = O(n^{-q/(q-1)})$. This implies that $\sum_{n=M+1}^{\infty} \rho_n$ can be made arbitrarily small by increasing $M$.

**Theorem 2 (Check-Regular Ensemble with Bounded Complexity).** Consider the ensemble of check-regular non-systematic IRA codes $\mathcal{C}$, where the d.d. of the parity-check nodes is given by

$$\rho(x) = x^2.$$

Assume that the transmission takes place over a BEC with erasure probability $p$, and let the d.d. of the information bit nodes be[4]

$$\lambda(x) = 1 + \frac{2p(1-x)^2 \sin\left(\frac{1}{3}\arcsin\left(\sqrt{-\frac{27p(1-x)^{\frac{3}{2}}}{4(1-p)^3}}\right)\right)}{\sqrt{3}\,(1-p)^4\left(-\frac{p(1-x)^{\frac{3}{2}}}{(1-p)^3}\right)^{\frac{3}{2}}}. \tag{8}$$

Let $\lambda_n$ be the coefficient of $x^{n-1}$ in the power series expansion of $\lambda(x)$ and, for an arbitrary $\epsilon \in (0,1)$, define $M(\varepsilon)$ to be the smallest positive integer[5] $M$ such that

$$\sum_{n=M+1}^{\infty} \frac{\lambda_n}{n} < \frac{(1-p)\varepsilon}{3}.$$

This infinite bit d.d. is truncated by treating all information bits with degree greater than $M(\varepsilon)$ as pilot bits (i.e., these information bits are set to zero). Let $\lambda_\varepsilon(x)$ be the $\varepsilon$-truncated d.d. of the bit nodes. Then, for all $p \in [0, 0.95]$, the polynomial $\lambda_\varepsilon(\cdot)$ has only non-negative coefficients, and the modified d.d. pair $(\lambda_\varepsilon, \rho)$ achieves a fraction $1-\varepsilon$ of the channel capacity with vanishing bit erasure probability under MPI decoding. Moreover, the complexity (per information bit) of encoding and decoding is *bounded* and satisfies

$$\chi_E(\varepsilon, \mathcal{C}) = \chi_D(\varepsilon, \mathcal{C}) < \frac{5}{(1-p)(1-\varepsilon)}.$$

In the limit as $\varepsilon$ tends to zero, the capacity is achieved with a *bounded* complexity of $\frac{5}{1-p}$.

The following two conjectures extend Theorems 1 and 2 to a wider range of parameters. Both of these conjectures can be proved by showing that the power series expansions of $\lambda(x)$ and $\rho(x)$ are non-negative for this wider range. Currently, we can show that the power series expansions of $\lambda(x)$ and $\rho(x)$ are non-negative over this wider range only for small values of $n$ (using numerical methods) and large values of $n$ (using asymptotic expansions). We note that if these conjectures hold, then Theorem 1 is extended to the range $p \in [0, \frac{3}{13}]$ (as $q \to \infty$), and Theorem 2 is extended to the entire range $p \in [0, 1)$.

**Conjecture 1.** The result of Theorem 1 also holds for $q \geq 4$ if

$$p \leq \begin{cases} \dfrac{6 - 7q + 2q^2}{6 - 13q + 8q^2} & 4 \leq q \leq 8 \\[4pt] \dfrac{12 - 17q + 6q^2}{12 - 37q + 26q^2} & q \geq 9 \end{cases}.$$

We note that Conjecture 1 is implied by the analysis in [7, Appendix A].

---

[4]For real numbers, one can simplify the expression of $\lambda(x)$ in (8). However, since we consider later $\lambda(\cdot)$ as a function of a complex argument, we prefer to leave it in the form of (8).

[5]The existence of $M(\varepsilon)$ for $\varepsilon \in (0,1)$ follows from the fact that $\lambda_n = O(n^{-3/2})$. This implies that $\sum_{n=M+1}^{\infty} \lambda_n/n$ can be made arbitrarily small by increasing $M$.

**Conjecture 2.** The result of Theorem 2 also holds for $p \in (0.95, 1)$.

In continuation to Theorem 2 and Conjecture 2, it is worth noting that Appendix C suggests a conceptual proof which in general could enable one to verify the non-negativity of the d.d. coefficients $\{\lambda_n\}$ for $p \in [0, 1-\varepsilon]$, where $\varepsilon > 0$ can be made arbitrarily small. This proof requires though to verify the positivity of a fixed number of the d.d. coefficients, where this number grows considerably as $\varepsilon$ tends to zero. We chose to verify it for all $n \in \mathbb{N}$ and $p \in [0, 0.95]$. We note that a direct numerical calculation of $\{\lambda_n\}$ for small to moderate values of $n$, and the asymptotic behavior of $\lambda_n$ (which is derived in Appendix B) strongly supports Conjecture 2.

**Theorem 3 (Information-Theoretic Bound on the Complexity of Punctured Codes over the BEC).** Let $\{\mathcal{C}'_m\}$ be a sequence of binary linear block codes, and let $\{\mathcal{C}_m\}$ be a sequence of codes which is constructed by randomly puncturing information bits from the codes in $\{\mathcal{C}'_m\}$.[6] Let $P_{\text{pct}}$ designate the puncturing rate of the information bits, and suppose that the communication of the punctured codes takes place over a BEC with erasure probability $p$, and that the sequence $\{\mathcal{C}_m\}$ achieves a fraction $1 - \varepsilon$ of the channel capacity with vanishing bit erasure probability. Then with probability 1 w.r.t. the random puncturing patterns, and for an arbitrary representation of the sequence of codes $\{\mathcal{C}'_m\}$ by Tanner graphs, the asymptotic decoding complexity under MPI decoding satisfies

$$\liminf_{m \to \infty} \chi_D(\mathcal{C}_m) \geq \frac{p}{1-p} \left( \frac{\ln\left(\frac{P_{\text{eff}}}{\varepsilon}\right)}{\ln\left(\frac{1}{1-P_{\text{eff}}}\right)} + l_{\min} \right)$$

where

$$P_{\text{eff}} \triangleq 1 - (1 - P_{\text{pct}})(1 - p)$$

and $l_{\min}$ designates the minimum number of edges which connect a parity-check node with the nodes of the parity bits.[7] Hence, a necessary condition for a sequence of randomly punctured codes $\{\mathcal{C}_m\}$ to achieve the capacity of the BEC with *bounded complexity* is that the puncturing rate of the information bits satisfies the condition $P_{\text{pct}} = 1 - O(\varepsilon)$.

Theorem 4 suggests an extension of Theorem 3, though as is clarified in [7], the lower bound in Theorem 3 is at least twice larger than the following lower bound when applied to the particular case of a BEC.

**Theorem 4 (Information-Theoretic Bound on the Complexity of Punctured Codes: General Case).** Let $\{\mathcal{C}'_m\}$ be a sequence of binary linear block codes, and let $\{\mathcal{C}_m\}$ be a sequence of codes which is constructed by randomly puncturing information bits from the codes in $\{\mathcal{C}'_m\}$. Let $P_{\text{pct}}$ designate the puncturing rate of the information bits, and suppose that the communication takes place over an MBIOS channel whose capacity is equal to $C$ bits per channel use. Assume that the sequence of punctured codes $\{\mathcal{C}_m\}$ achieves a fraction $1 - \varepsilon$ of the channel capacity with vanishing bit error probability. Then with probability 1 w.r.t. the random puncturing patterns, and for an arbitrary representation of the sequence of codes $\{\mathcal{C}'_m\}$ by Tanner graphs, the asymptotic complexity per iteration under MPI decoding satisfies

$$\liminf_{m \to \infty} \chi_D(\mathcal{C}_m) \geq \frac{1-C}{2C} \frac{\ln\left(\frac{1}{\varepsilon} \frac{1-(1-P_{\text{pct}})C}{2C \ln 2}\right)}{\ln\left(\frac{1}{(1-P_{\text{pct}})(1-2w)}\right)}$$

where

$$w \triangleq \frac{1}{2} \int_{-\infty}^{+\infty} \min\left(f(y), f(-y)\right) \, dy$$

and $f(y) \triangleq p(y|x=1)$ designates the conditional *pdf* of the channel, given the input is $x = 1$. Hence, a necessary condition for a sequence of randomly punctured codes $\{\mathcal{C}_m\}$ to achieve the capacity of an MBIOS channel with *bounded complexity per iteration* under MPI decoding is that the puncturing rate of the information bits satisfies $P_{\text{pct}} = 1 - O(\varepsilon)$.

---

[6]Since we do not require that the sequence of original codes $\{\mathcal{C}'_m\}$ is represented in a systematic form, then by saying 'information bits', we just refer to any set of bits in the code $\mathcal{C}'_m$ whose size is equal to the dimension of the code and whose corresponding columns in the parity-check matrix are linearly independent. This enables to define such a set in a non-unique way. If the sequence of the original codes $\{\mathcal{C}'_m\}$ is systematic (e.g., turbo or IRA codes before puncturing), then it is natural to define the information bits as the systematic bits of the code.

[7]The fact that the value of $l_{\min}$ can be changed according to the choice of the information bits is a consequence of the bounding technique.

**Remark 1 (Deterministic Puncturing).** It is worth noting that Theorems 3 and 4 both depend on the assumption that the set of information bits to be punctured is chosen randomly. It is an interesting open problem to derive information-theoretic bounds that apply to *every puncturing pattern* (including the best carefully designed puncturing pattern for a particular code). We also note that for any deterministic puncturing pattern which causes each parity-check to involve at least one punctured bit, the bounding technique which is used in the proofs of Theorems 3 and 4 becomes trivial and does not provide a meaningful lower bound on the complexity in terms of the gap (in rate) to capacity.

## 4   Code construction and Simulation Results

The performance of the check-regular construction (see Theorem 2) was evaluated using a number of codes. A fixed rate of 1/2 was chosen and non-systematic IRA codes of various block length and maximum degree were generated. For comparison, LDPC codes from the check-regular capacity-achieving ensemble [11] were constructed in the same manner. The IRA code ensembles were formed by truncating the algebraic d.d. function to $M = 25, 50$. The LDPC code ensembles were formed by choosing the check degree to be $d = 8, 9$ and then truncating the bit d.d. so that $\lambda(1) = 1$. This approach leads to maximum bit degrees of $M = 61, 126$, respectively. Actual codes of length $N = 8192, 65536$ and $524288$ were chosen from these ensembles, and simulated over the BEC. Computer simulation results are presented in Fig. 2 for a block length of $N = 8192, 65536$ and $524288$ bits. The examined ensembles are chosen randomly.

The random construction starts by quantizing the d.d. to integer values according to the block length. Next, it matches bit edges with check edges in a completely random fashion. Since this approach usually leads both multiple edges and 4-cycles, a post-processor is used. One iteration of post-processing randomly swaps all edges involved in a multiple-edge or 4-cycle events. This is repeated until there are no more multiple edges or 4-cycles. For the IRA codes, a single "dummy" bit is used to collect all of the edges originally destined for bits of degree greater than $M$. Since this bit is known to be zero, its column is removed to complete the construction. After this removal, the remaining IRA code is no longer check regular because this "dummy" bit is allowed to have multiple edges. In fact, it is exactly the check nodes which are reduced to degree 1 that allow decoding to get started. The bit-regular plot in Fig. 3 compares systematic IRA codes with $\lambda(x) = x^2$ and $\rho(x) = x^{36}$ (i.e., rate 0.925) with bit-regular non-systematic codes formed by our construction in Theorem 1 with $q = 3$. All of these codes do not use the outer Hamming code. Still, the error floors are much better because of all degree 3 bit nodes. Finally, a small number of systematic bits (100–200) are used to get decoding started. By comparing Theorems 1 and 2, we see that the bit-regular ensembles of IRA codes in Theorem 1 are limited to high rates (for $q = 3$, the rate should be at least $\frac{12}{13} \approx 0.9231$), in contrast to the check-regular ensembles (see Theorem 2 and Conjecture 2 in Section 3).

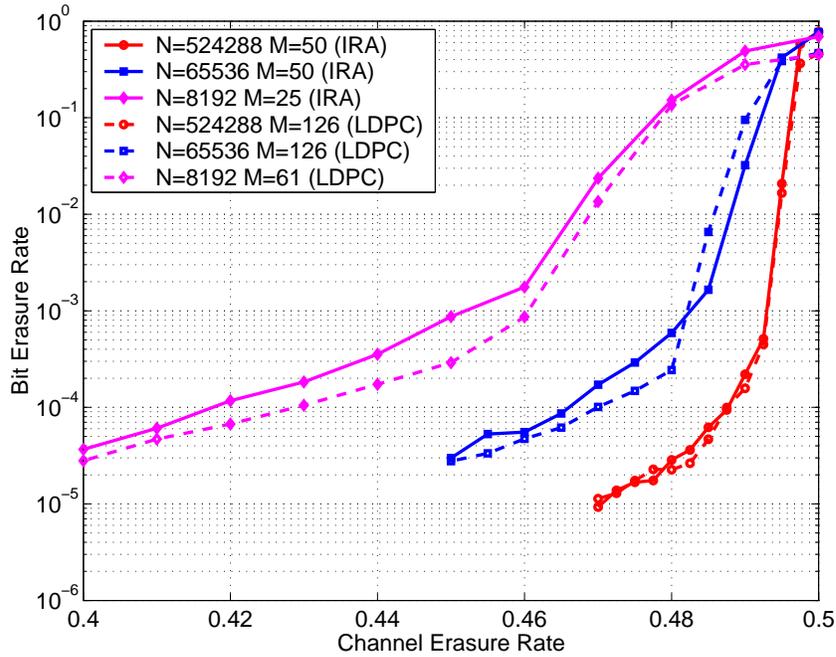

Figure 2: BER for random rate $1/2$ codes from the check-regular IRA ensemble in Theorem 2 and the check-regular LDPC ensemble [11] for $N = 8192, 65536$, and $524288$.

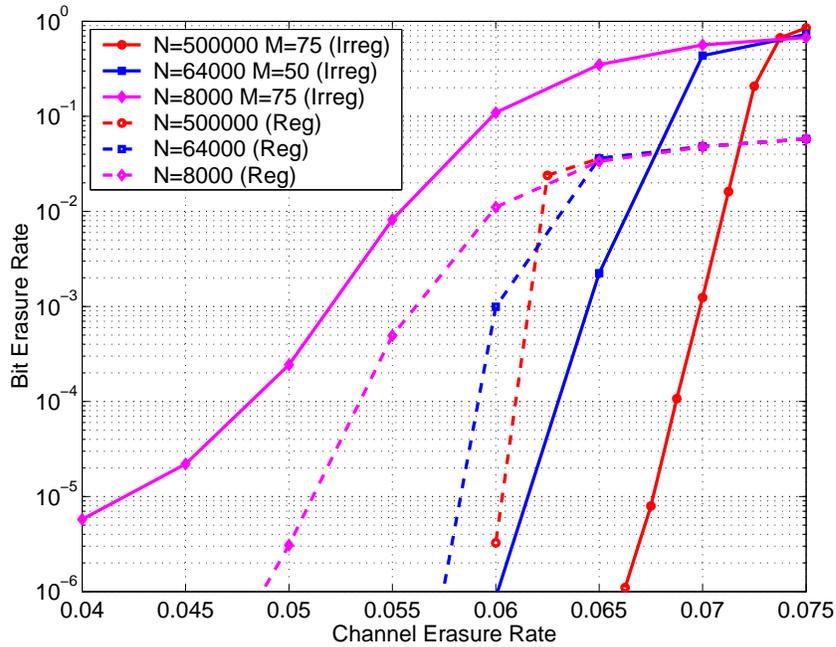

Figure 3: BER for random codes of rate $\frac{37}{40} = 0.9250$ from the bit-regular IRA ensemble in Theorem 1 with $q = 3$ and the ensemble of regular systematic IRA codes with d.d. $\lambda(x) = x^2$ and $\rho(x) = x^{36}$. The curves are shown for $N = 8000, 64000$, and $500000$.

# 5  Conclusions

In this work, we present two sequences of ensembles of non-systematic irregular repeat-accumulate (IRA) codes which asymptotically (as their block length tends to infinity) achieve capacity on the binary erasure channel (BEC) with *bounded complexity* (throughout this paper, the complexity is normalized per information bit). These are the first capacity-achieving ensembles with bounded complexity on the BEC to be reported in the literature. All previously reported capacity-achieving sequences have a complexity which grows at least like the log of the inverse of the gap (in rate) to capacity. This includes capacity-achieving ensembles of LDPC codes [3, 4, 11], systematic IRA codes [1, 10], and Raptor codes [12].

We show that under message-passing iterative (MPI) decoding, this new bounded complexity result is only possible because we allow a sufficient number of state nodes in the Tanner graph representing a code ensemble. The state nodes in the Tanner graph of the examined IRA ensembles are introduced by puncturing all the information bits. We also derive an information-theoretic lower bound on the decoding complexity of randomly punctured codes on graphs. The bound refers to MPI decoding, and it is valid for an arbitrary memoryless binary-input output-symmetric channel with a special refinement for the BEC. Since this bound holds with probability 1 w.r.t. a randomly chosen puncturing pattern, it remains an interesting open problem to derive information-theoretic bounds that can be applied to *every puncturing pattern*. Under MPI decoding and the random puncturing assumption, it follows from the information-theoretic bound that a necessary condition to achieve the capacity of the BEC with bounded complexity or to achieve the capacity of a general memoryless binary-input output-symmetric channel with bounded complexity per iteration is that the puncturing rate of the information bits goes to one. This is consistent with the fact that the capacity-achieving IRA code ensembles introduced in this paper are non-systematic, where all the information bits of these codes are punctured.

In Section 4, we use simulation results to compare the performance of our ensembles to the check-regular LDPC ensemble introduced by Shokrollahi [11] and to ensembles of systematic RA codes. For the cases tested, the performance of our check-regular IRA codes is slightly worse than that of the check-regular LDPC codes. It is clear from these results that the fact that these capacity-achieving ensembles have bounded complexity does not imply that their performance, for small to moderate block lengths, is superior to other reported capacity-achieving ensembles. Note that for *fixed* complexity, the new codes eventually (for $n$ large enough) outperform any code proposed to date. On the other hand, the *convergence speed* to the ultimate performance limit happens to be quite slow, so for small to moderate block lengths, the new codes are not necessarily record breaking. Further research into the construction of codes with bounded complexity is likely to produce codes with better performance for small to moderate block lengths.

The central point in this paper is that by allowing state nodes in the Tanner graph, one may obtain a significantly better tradeoff between performance and complexity as the gap to capacity vanishes. Hence, it answers in the affirmative a fundamental question which was posed in [9] regarding the impact of state nodes (or in general, more complicated graphical models than bipartite graphs) on the performance versus complexity tradeoff under MPI decoding. Even the more complex graphical models, employed by systematic IRA codes, provides no asymptotic advantage over codes which are presented by bipartite graphs under MPI decoding (see [9, Theorems 1, 2] and [10, Theorems 1, 2]). Non-systematic IRA codes do provide, however, this advantage over systematic IRA codes; this is because the complexity of systematic IRA codes becomes unbounded, under MPI decoding, as the gap to capacity goes to zero.